\documentclass[aps,pra,twocolumn,groupedaddress,showpacs,superscriptaddress,amssymb,amsmath]{revtex4-2}
\usepackage{pgfplots}
\pgfplotsset{compat=1.13}
\usepackage[utf8]{inputenc}
\usepackage{graphicx}
\usepackage{subcaption}
\usepackage{tabularx}
\usepackage{xcolor}
\usepackage{amsmath}

\usepackage{comment}
\usepackage{dcolumn}
\usepackage{hyperref}
\hypersetup{
	colorlinks=true,
	linkcolor=blue,
	citecolor=cyan
}
\usepackage{bm}
\usepackage{epsf}
\usepackage{braket}
\usepackage{tensor}
\usepackage{soul}
\usepackage{float}
\usepackage{mathrsfs}
\usepackage{mathtools}
\usepackage{multirow}
\usepackage{caption}
\usepackage{amsfonts}

\newcommand{\be}{\begin{equation}}
	\newcommand{\ee}{\end{equation}}
\newcommand{\bea}{\begin{eqnarray}}
	\newcommand{\eea}{\end{eqnarray}}

\makeatletter
\newsavebox{\@brx}
\newcommand{\llangle}[1][]{\savebox{\@brx}{\(\m@th{#1\langle}\)}%
	\mathopen{\copy\@brx\kern-0.5\wd\@brx\usebox{\@brx}}}
\newcommand{\rrangle}[1][]{\savebox{\@brx}{\(\m@th{#1\rangle}\)}%
	\mathclose{\copy\@brx\kern-0.5\wd\@brx\usebox{\@brx}}}
\makeatother

\begin{document}
\title{Steady state correlation function beyond the standard weak coupling limit and consistency with  KMS relation }
\author{Sakil Khan}
\email{sakil.khan@students.iiserpune.ac.in}
\affiliation{Department of Physics,
		Indian Institute of Science Education and Research, Pune 411008, India}
\author{Lokendra Singh Rathore}
\email{lokendra.rathore@students.iiserpune.ac.in}
\affiliation{Department of Physics,
		Indian Institute of Science Education and Research, Pune 411008, India}
\author{ Sachin Jain  }
\email{sachin.jain@iiserpune.ac.in}
\affiliation{Department of Physics,
		Indian Institute of Science Education and Research, Pune 411008, India}
\date{\today}
\begin{abstract}
 Thermalization of a system when interacting with a thermal bath is an interesting problem. If a system eventually reaches a thermal state in the long time limit, it's expected that its density matrix would resemble the mean-force Gibbs state. Moreover, the correlation function must satisfy the Kubo-Martin-Schwinger (KMS) condition or equivalently the Fluctuation-Dissipation Relation (FDR). In this paper, we derive a formal expression for the non-Markovian two-point function within the context of the weak coupling limit. Using this expression, we explicitly compute the two-point function for specific models, demonstrating their adherence to the KMS. In addition, we have formulated a  non-perturbative approach in the form of a self-consistent approximation that includes a partial resummation of perturbation theory. This approach can capture strong coupling phenomena while still relying on simple equations. Notably, we verify that the two-point function obtained through this method also satisfies the KMS condition.

\end{abstract}
  \maketitle  


\section{Introduction}
In recent times, there has been a growing interest in understanding the conditions and mechanisms governing the thermalization of a small (microscopic) open quantum system \cite{breuer,Carmichael,schaller2014open,xiong2020exact,PhysRevB.107.125149,PhysRevE.100.022111,PhysRevB.97.134301,PhysRevB.86.155424,PhysRevLett.121.170402}. If the system's degrees of freedom are negligible compared to the bath degrees of freedom, one would naively expect that the system will go to a steady state at the long time limit.
However, before addressing the question of whether the stationary state is thermal or not, it is
necessary to identify criteria that allow a clear-cut detection of
thermodynamic equilibrium conditions in the stationary state.

In this regard, it is important to consider not only the static properties of the density matrix of the system, which describes its stationary state but also the dynamics of fluctuations. These fluctuations are encoded, for example, in the two-time correlation function (adhering to the KMS relation). More precisely, at the long time limit, (a) the density matrix of the system should be the mean-force density matrix \cite{PhysRevLett.127.250601,PhysRevLett.129.200403},
and (b) the correlation function of the system at the long time must satisfy the KMS condition or equivalently the Fluctuation-Dissipation Relation (FDR) \cite{kubo1957statistical, MS, breuer,Carmichael}.

 A lot of work has already been done to address the first issue \cite{PhysRevLett.127.250601,PhysRevLett.129.200403}. However, till now, the discussion about the second issue is very limited in the literature.
 To fill this gap, in this paper, our central focus will be to answer this question as clearly as possible.

Except for exactly solvable models like Caldeira-Leggett \cite{CALDEIRA1983587}, we generally resort to approximate techniques to explicitly compute the correlation function. In this context, it remains unclear whether the approximate correlation function adheres to the FDR, or, in other words, which approximation may violate it. Addressing this, L. Sieberer et al.\cite{sieberer2015thermodynamic,sieberer2016keldysh} recently discussed the constraints and symmetries imposed on the total Swinger-Keldysh (SK) action or, equivalently, on the total system-bath Hamiltonian. These constraints guide the system toward thermalization at the long time limit.

In the first part of our paper, we employed an approximate technique called the image operator method \cite{Sakil1,karve2020heisenberg,khan2023modified}. We use this method to calculate the two-point correlation function, and then we explicitly demonstrate that the correlation function satisfies the KMS relation. However, it's important to note that the applicability of this method is limited to specific examples; it does not universally guarantee the satisfaction of the KMS relation for a generic system. Furthermore, the validity of this method is contingent upon the assumption of weak system-bath coupling.

To go beyond the standard weak coupling limit, in the next part of our paper, we have developed a self-consistent non-perturbative technique\cite{haertle2013decoherence, erpenbeck2021revealing, pruschke1993hubbard, meir1993low,chen2016anderson, eckstein2010nonequilibrium, nordlander1999long, muller1984self, bickers1987review,ganguly2023study}, namely the self-consistent Born approximation \cite{altland2010condensed,bruus2004many} or the NCA approach, following the Swinger-Keldysh path integral. Using this technique, we abstractly show that the correlation function must obey the FDR relation for a general class of systems. The FDR follows from the KMS relation \cite{chaudhuri2019spectral,haehl2017thermal}. More precisely, the FDR is equivalent to a combination of quantum mechanical time reversal and the KMS condition \cite{sieberer2015thermodynamic}.

We organize the paper as follows:  In section \ref{sec-II}, we derived a formal expression of the two-point function using the image operator method. In section \ref{sec-III}, we explain the NCA technique and show that the two-point function obtained using this technique will always respect the FDR. In section \ref{sec-IV}, we compare the Green's function obtained using the Born and Self-consistent Born approximations. We then summarize our work in section \ref{sec-V}. In Appendix \ref{appa}, we provide a detailed derivation of the two-point function using the image operator method. In Appendix \ref{appb}, we determine the Laplace transformation of a reduced system operator. In Appendix \ref{appc}, we present the explicit form of the Schwinger Keldysh Action for the general class of systems addressed in this paper. In Appendix \ref{lblb10}, we establish that if the self-energy satisfies the FDR, the Green's function related to it by the Dyson equation also satisfies the FDR. Appendix \ref{lblb11} demonstrates a condition under which the leading-order self-energy term satisfies the FDR. In Appendix \ref{lblb13}, we derive analytical expressions for steady-state Green's functions compared in Section \ref{sec-IV} and then verify the FDR for them.

\section{Steady state correlation function under the weak coupling limit}\label{sec-II}
In this section, we want to derive the two-point correlation function at the steady state up to the leading order in system-bath coupling strength. For our setup,  we take the following general form of the total Hamiltonian 
\begin{equation} \label{eq1nma}
		\begin{split}
			H & = H_{S}+H_{R}+ H_{S  R} \\
			& = H_{S}+\sum_{k} \Omega_{k} b_{k}^\dagger b_{k}+ \sum_{k} \alpha_{k}( S b_{k}^\dagger+ S^\dagger   b_{k})\;,
		\end{split}
\end{equation}
where $b_{k} $ and $b_{k}^\dagger $ represent the bosonic or fermionic annihilation and creation operator for the $k$-th mode, respectively. The third term in Eq.\eqref{eq1nma} represents the system-bath coupling with the generic system operator $S$ coupled with the $k$-th bath mode with interaction strength $\alpha_k$. 

Our first aim is to calculate the two-point correlation function of the form 
\begin{equation}
  \langle O_{1}(t+\tau)O_{2}(t) \rangle={\rm Tr}_{S}\Big[	[O_{1  }(t+\tau)O_{2}(t)]_{S} \,\rho_{S}(0)\Big] ,
\end{equation}
where $	[O_{1  }(t+\tau)O_{2}(t)]_{S}$
denotes the two-point reduced operator \cite{Sakil1,karve2020heisenberg} defined as:
\begin{align}
    [O_{1  }(t+\tau)O_{2}(t)]_{S} = {\rm Tr}_{B} \big[ O_{1}(t +\tau) O_{2}(t) \, \rho_{B} \big]
\end{align}
We follow the recipe of Ref.~\cite{karve2020heisenberg}, to express the two-point reduced operator $	[O_{1  }(t+\tau)O_{2}(t)]_{S}$, up to the leading order in the system-bath coupling, in terms of one-point reduced operators ($O_{1 S }$ and $O_{2 S }$) (for details see Appendix-\ref{appa}). Note that, the dynamics of the reduced one-point operator is governed by the Born master equation \cite{breuer}.

For the system defined in Eq.\eqref{eq1nma}, we can explicitly show the two-point reduced operator is given by 
\begin{widetext}
    \begin{equation} \label{eq2ptst}
	[O_{1  }(t+\tau)O_{2}(t)]_{S}=O_{1 S}(t+\tau)O_{2 S}(t)+ D_{1}(t,\tau)+D_{2}(t,\tau)\;,
\end{equation}
In Appendix-\ref{appa}, we have given the details of this derivation. Note that, to derive the above expression we have only considered the weak coupling approximation i.e., we keep terms up to the leading order in the system-bath coupling. The two-point reduced operator is not just the product of one-point reduced operators, we also get two  inhomogeneous  terms $D_{1}(t, \tau)$  and  $D_{2}(t, \tau)$ which are given by the following equations
\begin{align}\label{d1}
	D_{1}(t, \tau)\!=\!\!\!\!\!\sum_{j,l,m,n}\!
\int\! \frac{ d\Omega}{2\pi}  F_{\eta}(\Omega)\;e^{i\Omega \tau} e^{-i(\omega_{j}+\omega'_{m})(t+\tau)}
 e^{-i(\tilde{\omega}_{l}-\omega'_{n})t}\!\!\!\int_{0}^{t+\tau} \!\!\!\!\!d\tau'_{1}\!\Big[ O_{1 j S}(\tau'_{1}),
	S_{m}\Big] e^{-i[\Omega-(\omega_{j}+\omega'_{m})]\tau'_{1}}\!\!\!\int_{0}^{t}\!\!\!d\tau'_{2}\Big [ S^\dagger_{n}, O_{2 l  S}(\tau'_{2})
	\Big]  e^{i[\Omega+(\tilde{\omega}_{l}-\omega'_{n})]\tau'_{2}}\;,
\end{align}
\begin{align}\label{d2}
	D_{2}(t, \tau)\!=\!\!\!\!\! \sum_{j,l,m,n}
\!\int\! \frac{ d\Omega}{2\pi} \tilde  {F}_{\eta}(\Omega)e^{-i\Omega \tau} e^{i(\omega'_{m}\!-\omega_{j})(t+\tau)} e^{-i(\tilde{\omega}_{l}+\omega'_{n})t}\!\!\! \int_{0}^{t+\tau} \!\!\!\!d\tau'_{1}\!\Big[ O_{1 j S}(\tau'_{1}),
	S^\dagger_{m}\Big]e^{i[\Omega+(\omega_{j}-\omega'_{m})]\tau'_{1}}
\!\!\! \int_{0}^{t}\!\!d\tau'_{2}\Big [ S_{n}, O_{2 l  S}(\tau'_{2})
	\Big]  e^{i[(\tilde{\omega}_{l}+\omega'_{n})-\Omega]\tau'_{2}}
\end{align}
 \end{widetext}
 where $F_{\eta}(\Omega)=J(\Omega) n_{\eta}(\Omega)$ and  $\tilde{F}_{\eta}(\Omega)=(J(\Omega)\!-\!\eta F_{\eta}(\Omega)) $ with $J(\Omega)$ is the spectral density function of the bath which is defined as $J(\Omega)= 2\pi \sum_{k} |\alpha_{k}|^2 \delta(\Omega-\Omega_{k})$. Note that, here  $  n_{\eta}(\Omega)$ represents the Bose or Fermi distribution function i.e. $F_{\eta}(\Omega)=[ e^{\beta \Omega}+ \eta]^{-1}$ with $\eta=+1$ and $\eta=-1$ are for fermions and bosons, respectively. $\omega_j$ corresponds to the possible energy differences between the bare system eigenenergies that appear by performing spectral decomposition for the operator $O_{1 S}$. In other words, we use the fact that 
\begin{equation}\label{specttral1}
O_{1 S}(t)=\sum_{j}  O_{1 j  S}(t-t') e^{-i\omega_{j}t'}+O(\alpha_{k}).
\end{equation}
Similarly, $\tilde{\omega}_l$ and $\omega_m'$ correspond to the possible energy differences of the bare system for the operators $O_{2 S}$ and $S$ respectively. Eq.\eqref{eq2ptst}-Eq.\eqref{d2} represents the two-point correlation function at any time.

For the rest of the paper, we focus on the steady-state correlation function. To obtain the steady state correlation function, we simply need to take the $t \to \infty$
limit in Eq.\eqref{eq2ptst}-Eq.\eqref{d2}. Below, we will illustrate the calculation of the steady state correlation function using the image operator method for two paradigmatic models. Additionally, we will explicitly show that the obtained correlation function satisfies the KMS condition.

\section**{\textbf{Two-point correlation function at the steady state and KMS for some specific models:}}\\

Here, using the expression obtained in Eq.\eqref{eq2ptst}-Eq.\eqref{d2}, we compute the correlation function for a dissipative non-interacting bosonic/fermion system and for the dissipative spin-boson model. For both these models, we show explicitly that the correlation function satisfies the KMS condition in the long-time limit. 
\subsection*{\it Dissipative Bosonic/Fermionic Model}
We treat a single bosonic or fermionic degree of freedom as a system that is coupled to a corresponding bosonic or fermionic thermal bath. The total Hamiltonian is given by
\begin{equation} \label{CL}
		H
		= \omega_{0} a^\dagger a+\sum_{k} \Omega_{k} b_{k}^\dagger b_{k}+ \sum_{k} \alpha_{k}( a\, b_{k}^\dagger+a^\dagger \, b_{k}),
\end{equation}
where $a $ and $a^\dagger $ represent the bosonic or fermionic annihilation and creation operator for the system, respectively.
For this model, we want to calculate $\langle a^\dagger(t+\tau)a(t)\rangle$  up to the leading order in system-bath coupling at the steady state using Eq.\eqref{eq2ptst}-Eq.\eqref{d2}.  Let us first note that, for this correlation function, $O_{1 }=a^\dagger$ and $O_{2}=a$ in Eq. \eqref{eq2ptst}. With this identification, it is easy to show that the $D_{1}$ term (expressed in Eq.\eqref{d1}) at the long time limit i.e. $t \to \infty$, takes this interesting form,
 \begin{align} \label{D1}
 D_{1}&= \int \frac{ d\Omega}{2\pi}  F_{\eta}(\Omega)\;e^{i\Omega \tau}\int_{0}^{\infty} \!d\tau'_{1} \Big[ a^\dagger_{S}(\tau'_{1}),
	a \Big] e^{-i\Omega\tau'_{1}}\nonumber\\
 & \;\;\;\;\;\;\;\;\;\;\;\;\;\;\;\;\;\;\int_{0}^{\infty} d\tau'_{2}\Big [ a^\dagger, a_{ S}(\tau'_{2})	\Big]e^{i\Omega\tau'_{2}}\nonumber\\
 & = \int \frac{ d\Omega}{2\pi}  F_{\eta}(\Omega)\;e^{i\Omega \tau} \Big[ \tilde{a}^\dagger_{S}(i\Omega),
	a \Big] 
\Big [ a^\dagger, \tilde{a}_{ S}(-i\Omega)	\Big]\;,
\end{align}
where $\tilde{a}_{ S}(-i\Omega)$ is the Laplace transformation of $a_{ S}(t)$ i.e. $\tilde{a}_{ S}(-i\Omega)= \int_{0}^{\infty} dt\;  a_{ S}(t)	e^{i\Omega t}$. It is interesting to note that the Laplace transform is a feature specific to the steady state and appears naturally in the steady state limit.
Similarly, we can show that $D_{2}$ (expressed in Eq. \eqref{d2}) is zero for this model. Note that for this model, the first term of Eq.\eqref{eq2ptst} is zero since $ a_{ S}(\infty)= a^\dagger_{ S}(\infty)=0$ and  the Laplace transformation of $a_{ S}(t) $ is (see Appendix-\ref{appb} for details)
\begin{equation}
     \tilde{a}_{ S}(-i\Omega)=\frac{a}{\big[-i \big(\Omega-\omega_{0}-\Sigma(\Omega)\big)+ \frac{J(\Omega)}{2}\big]} .
\end{equation}
Making these substitutions in Eq.\eqref{D1}, we get the following expression for the two-point correlation function at the steady state 
\begin{align}\label{ada}
\langle a^\dagger(t+\tau)a(t)\rangle_{\text{ss}}\!= \!\!\!\int^{\infty}_{0
  } \!\frac{ d\Omega}{2\pi} \frac{e^{i\Omega \tau} F_{\eta}(\Omega)} {\!\Big[\!\big(\Omega-\omega_{0}-\Sigma(\Omega)\big)^2\!\!+\!\!\big(J(\Omega)/2\big)^2\Big]}\;.
\end{align}
It is worth noting how the two-point correlator expressed by the seemingly complicated Eq.\eqref{eq2ptst}-Eq.\eqref{d2}, takes on an elegant and simple form in the steady state limit. Another important feature of the above equation is that at $\tau = 0$, the right-hand side becomes the steady-state one-point expectation value $\langle a^{\dagger} a \rangle_{ss}$. We want to determine whether the correlation function obtained in Eq.\eqref{ada} satisfies the KMS condition or not. The KMS condition states that:
\begin{align}
    \langle a^\dagger(t+\tau)a(t)\rangle_{\text{ss}} = \langle a(t)a^\dagger(t+\tau - i \beta)\rangle_{\text{ss}}
\end{align}
So, to check the KMS condition, we need to find the other correlator also, i.e., $\langle a(t) a^\dagger(t+\tau) \rangle$. Following the exact similar steps, one can easily compute the following correlator
\textcolor{black}{\begin{align} \label{eqaad}
		\langle a(t) a^\dagger(t+\tau) \rangle_{\text{ss}} \!=\!\!\!\int^{\infty}_{0
  } \!\frac{ d\Omega}{2\pi}\frac{e^{i\Omega \tau}  \big (1 - \eta \, n_{\eta}(\Omega)\big)J(\Omega)} {\Big[\!\big(\Omega\!-\omega_{0}\!-\!\Sigma(\Omega)\big)^2\!\!+\!\!\big(J(\Omega)/2\big)^2\Big]} 
\end{align}}
Looking at Eq.\eqref{ada} and Eq.\eqref{eqaad}, it is clear that 
the above correlators satisfy the KMS condition at the steady state.
 Moreover, the obtained expression for the correlation functions in Eq.\eqref{ada} and Eq.\eqref{eqaad} turns out to be the exact correlation function at the steady state. 
 
\subsection*{\it Dissipative  Spin-Boson Model (Secular)}
We want to apply the same method to calculate the correlation function for another paradigmatic model, namely the dissipative spin-boson model. The total Hamiltonian of the system is given by
\begin{equation} \label{CL}
		H
		= \frac{\omega_{0}}{2} \sigma_{z}+\sum_{k} \Omega_{k} b_{k}^\dagger b_{k}+ \sum_{k} \alpha_{k}( \sigma_{-}\, b_{k}^\dagger+\sigma_{+} \, b_{k})\;.
\end{equation}
where $b_{k} $($b^\dagger_{k} $) represents the bosonic annihilation (creation) operator and $\sigma_{-}$ ($\sigma_{+}$) is the lowering (raising) operator of the spin-half system. Our aim is to compute  $\langle \sigma_{+}(t+\tau)\sigma_{-}(t)\rangle$  up to the leading order in the system-bath coupling.  Let us first note that, $O_{1 }=\sigma_{+}$ and $O_{2}=\sigma_{-}$ in  Eq. \eqref{eq2ptst}. With this identification, it is easy to show that the $D_{1}$ term (expressed in Eq.\eqref{d1}) at the long time limit i.e. $t \to \infty$, is given by
 \begin{align} \label{d1s}
 D_{1}&= \int \frac{ d\Omega}{2\pi}  F_{-}(\Omega)\;e^{i\Omega \tau}\int_{0}^{\infty} \!d\tau'_{1} \Big[ \sigma_{+ S}(\tau'_{1}),
	\sigma_{-} \Big] e^{-i\Omega\tau'_{1}}
\nonumber\\
&\;\;\;\;\;\;\;\;\;\;\;\;\;\;\;\;\;\int_{0}^{\infty} d\tau'_{2}\Big [ \sigma_{+}, a_{- S}(\tau'_{2})	\Big]e^{i\Omega\tau'_{2}}\nonumber\\
 & = \int \frac{ d\Omega}{2\pi}  F_{-}(\Omega)\;e^{i\Omega \tau} \Big[ \tilde{\sigma}_{+ S}(i\Omega),
	\sigma_{-} \Big] 
\Big [ \sigma_{+}, \tilde{\sigma}_{ - S}(-i\Omega)	\Big]\;,
\end{align}
where $\tilde{\sigma}_{ - S}(-i\Omega)$ is the Laplace transformation of $\sigma_{- S}(t)$ i.e. $\tilde{\sigma}_{- S}(-i\Omega)= \int_{0}^{\infty} dt\;  \sigma_{- S}(t)	e^{i\Omega t}$.
Similarly, we can show that $D_{2}$ (expressed in Eq.\eqref{d2}) is zero for this model. Finally, by substituting the Laplace transformation of the operators appearing in Eq. \eqref{d1s}, we get the simple and explicit form of the two-point correlation function at the steady-state 
 \begin{align}\label{spm}
&\langle \sigma_{+}(t+\tau)\sigma_{-}(t)\rangle_{\text{ss}}
 \nonumber\\
 &= \int^{\infty}_{0
  } \!\frac{ d\Omega}{2\pi}\frac{e^{i\Omega \tau} F_{-}(\Omega)} {\!\Big[\!\big(\Omega\!-\omega_{0}\!-\!\Sigma''(\Omega)\big)^2\!\!+\!\!\big((n(\Omega)\!\!+\!\!1/2)J(\Omega)\big)^2\Big]}\;.
\end{align}
By following the identical  steps, we can find the following correlator
\begin{align}\label{smp}
	&\langle \sigma_{-}(t+\tau)\sigma_{+}(t)\rangle_{\text{ss}}
	\nonumber\\
	&= \int^{\infty}_{0
	} \!\frac{ d\Omega}{2\pi}\frac{e^{-i\Omega \tau} (1+n_{-}(\Omega))J(\Omega)} {\!\Big[\!\big(\Omega\!-\omega_{0}\!-\!\Sigma''(\Omega)\big)^2\!\!+\!\!\big((n(\Omega)\!\!+\!\!1/2)J(\Omega)\big)^2\Big]}\;.
\end{align}
It is evident from  Eq.\eqref{spm} and Eq.\eqref{smp} that 
the above correlators satisfy the KMS condition at the steady state.
Note that, the above expressions of the correlation function are obtained under only the weak coupling limit. 

Note that, by employing the image operator method developed in this section, we explicitly calculate the two-point correlation function at the steady state for specific examples and demonstrate that they satisfy the KMS relation. Our next aim is to develop a technique for calculating the correlation function that ensures consistency with the KMS condition for a generic model. To be more precise, our technique will provide insight into the KMS relation without explicit computation of the correlation function, Additionally, we aim to go beyond the standard weak coupling limit for calculating correlation functions.

To achieve our goal, in the next section, we have developed an approximate and self-consistent non-perturbative technique using the Swinger-Keldysh path integral. This technique not only enables us to comment on the KMS relation for a generic type of system but also allows us to go beyond the weak coupling limit.

\section{Correlation Functions beyond the Standard Weak Coupling Limit}\label{sec-III}

In this section, we initially develop a framework that will enable us to extend beyond the standard weak coupling limit for calculating the steady-state two-point function. We achieve this by applying the ``self-consistent Born approximation" (or Non-Crossing Approximation, NCA). Then, we proceed to demonstrate that this approximation, while providing a simple method for calculating the two-point function, also ensures that the two-point correlators obey the Fluctuation-Dissipation Relation (FDR).

For this, we focus on a very general class of open quantum systems comprising a single bosonic mode coupled to a Gaussian bosonic bath, with the total Hamiltonian $H = \ H_{S} + H_{B} + H_{I,l} + H_{I,nl}$, where we take $H_{S}$ to be an arbitrary system Hamiltonian, and
\begin{align} \label{lbl1}
    &H_{B} = \sum_{k} \omega_{k} b^{\dagger}_{k} b_{k}, \ \ \
    H_{I,l} = \sum_{k} \alpha_{k} (a^{\dagger}b_{k} + a b^{\dagger}_{k}) ,\nonumber \\  
    &H_{I,nl} = \sum_{k} \alpha_{k} ( a^{\dagger m} a^{n} b_{k} + a^{\dagger n} a^{m} b^{\dagger}_{k}).
\end{align}
Here, $b_{k}$ and $b_{k}^{\dagger}$ represent the bosonic annihilation and creation operators for the k-th bath mode, respectively. The third term, $H_{I,l}$ in Eq.\eqref{lbl1}, signifies the system-bath coupling through linear operators of the system, coupled to the k-th bath mode with an interaction strength of $\alpha_{k}$. The fourth term, $H_{I,nl}$ represents the system-bath coupling employing a generic non-linear operator ($m, n \ge 0$), coupled to the k-th bath mode with an interaction strength of $\alpha_{k}$. Note that the non-linear system operator has been considered to be normally ordered in the second-quantization notation.

Our objective here is to study the steady-state green's function for the system. To accomplish this, we promote the total Hamiltonian to the Schwinger Keldysh Path Integral. The resulting Schwinger Keldysh Action is expressed in Eq. \eqref{lbl16} of Appendix-\ref{appc}, in the classical-quantum (cl-q) basis for the field operators. The Schwinger-Keldysh functional integral, involving both system and bath degrees of freedom in the total action, is quadratic in the bath degrees field.  Assuming that the bath is in a thermal state, we can integrate it out, resulting in an action expressed solely in system degrees of freedom. This action takes the form $S = S_{S} + S'_{l} + S'_{nl}$, as expressed in Eq.\eqref{lbl17}.

We employ the standard tool of Feynman diagrams to perturbatively calculate the self-energy `$\tilde{\Sigma}(\omega)$' for the steady-state Green's function. The Dyson series
\begin{align} \label{lbl2}
    G(\omega) &= G_{(0))}(\omega) + G_{(0)}(\omega) \tilde{\Sigma}(\omega) G_{(0)}(\omega) + ... \nonumber \\
    &= G_{(0)} + G_{(0)}\tilde{\Sigma}(\omega) G  =  \bigl(G^{-1}_{(0)}(\omega) - \tilde{\Sigma}(\omega)\bigr)^{-1},
\end{align}
relates the self-energy to the Green's function, where $G_{(0)}(\omega)$ represents the system's bare Green's function. For the class of models described by the Hamiltonian in Eq.\eqref{lbl1}, the self-energy always takes the form:
\begin{align} \label{lbl3}
    \tilde{\Sigma} = \tilde{\Sigma}_{S} + \tilde{\Sigma}_{I,l} + \tilde{\Sigma}_{I,nl} .
\end{align}
Here,  $\tilde{\Sigma}_{S}$, $\tilde{\Sigma}_{I,l}$, and $\tilde{\Sigma}_{I,nl}$ represent contributions to the self-energy arising from $S_{S}$, $S'_{l}$, and $S'_{nl}$, and correspondingly from $H_{S}$, $H_{I,l}$, and $H_{I,nl}$, respectively.

  The `standard Born approximation' involves expressing the Dyson series up to the leading order in the self-energy. In contrast, the self-consistent Born approximation constitutes a straightforward non-perturbative method that involves a partial resummation of perturbation theory. The self-consistent approximations are widely used in various contexts in many-body physics, such as quantum transport \cite{haertle2013decoherence, erpenbeck2021revealing, pruschke1993hubbard, meir1993low} problems and quantum impurity models, both in and out of equilibrium \cite{chen2016anderson, eckstein2010nonequilibrium, nordlander1999long, muller1984self, bickers1987review}. Moreover, it naturally appears in the physics of large-N systems \cite{altland2010condensed, bruus2004many}. This approach is capable of capturing strong coupling phenomena while still relying on simple equations \cite{ganguly2023study}. It is also referred to as the `Non-Crossing Approximation' because, heuristically, it can also be implemented by considering all orders of Feynman diagrams, which can be drawn on the plane in such a way that the propagators don't cross each other and intersect only at the vertices. FIG. \ref{fig1} depicts such diagrams for a simple example.
 
 Implementing the self-consistent Born approximation involves two steps:
\begin{enumerate}
    \item Write down the Dyson series with the Born approximation for the self-energy, \label{lb1}
    \item Replace the `bare Green's function' inside self-energy with the total Green's function. \label{lb2}
\end{enumerate} 

To demonstrate how to implement the self-consistent Born approximation, we consider the total Hamiltonian of the form given by Eq.\eqref{lbl1}, with the simplest non-linear interaction Hamiltonian $H_{I,nl}$:
\begin{align}\label{lbl4}
    H_{I,nl} = \sum_{k} \alpha_{k} a^{\dagger} a(b^{\dagger}_{k} + b_{k})
\end{align}
\textit{Step} \ref{lb1}: First, we calculate the self-energy at the leading order. Its form is expressed by Eq.(\ref{lbl3}). 
The contribution to the leading-order self-energy from the system Hamiltonian $H_{S}$, and interaction Hamiltonians 
$H_{I,l}$ and $H_{I,nl}$ respectively are:
\begin{align}
    \tilde{\Sigma}_{S}^{(2)}(\omega) = \begin{pmatrix}
        0 & \omega_{0}  \\ \omega_{0} & 0
    \end{pmatrix},
\end{align}
\begin{align} \label{lbl5}
    \tilde{\Sigma}_{I,l}^{(2)}(\omega) =
    \begin{pmatrix}
        0 & i J(\omega) + \Sigma(\omega) \\ - i J(\omega) + \Sigma(\omega) & -2i J(\omega) \coth(\frac{\beta}{2}\omega) 
    \end{pmatrix} ,
\end{align}
\begin{align} \label{lbl6}
    \tilde{\Sigma}^{(2)}_{I,nl}(\omega) = \int_{\tau} D_{B}(\tau) \circ G_{(0)}(\tau) e^{i \omega \tau} .
\end{align}
Here, $J(\omega)$ represents the bath spectral density function, and $\Sigma(\omega) = \mathcal{P}\int_{\omega'} \frac{1}{\pi}\frac{J(\omega')}{\omega - \omega'}$. Here, `$D_{B}$' denotes the bath Green's function. The symbol `$\circ$' is a concise representation of the Feynman diagrams arising from standard Wick contractions. Its definition is provided in Appendix \ref{lblb11}. The corresponding Dyson series gives the steady state Green's function under Born approximation:
\begin{align} \label{lbl14}
    G_{B}(\omega) = \bigl(G^{-1}_{(0)}(\omega) - \tilde{\Sigma}_{S}^{(2)}(\omega) -  \tilde{\Sigma}_{I,l}^{(2)}(\omega) - \tilde{\Sigma}_{I,nl}^{(2)}(\omega)\bigr)^{-1}.
\end{align}

\textit{Step} \ref{lb2}: Now, to implement the self-consistent Born approximation, we replace the bare Green's function in the above self-energy with the total Green's function. For our example, the self-energy $\tilde{\Sigma}_{S}^{(2)}(\omega)$ is independent of $G_{(0)}$. So, it remains unchanged under this approximation. Similarly, the self-energy $\tilde{\Sigma}_{I,l}$ and is always independent of $G_{(0)}$. So, it always remains unchanged under this approximation. However, the self-energy contribution $\tilde{\Sigma}_{I,nl}^{(2)}$, as expressed in Eq.\eqref{lbl4}, changes to:
\begin{align} \label{lbl7}
    \tilde{\Sigma}_{I,nl}^{SC}(\omega) =   \int_{\tau} D_{B}(\tau) \circ G^{SC}(\tau) e^{i \omega \tau} ,
\end{align}
giving the self-consistent steady-state Green's function
\begin{align} \label{lbl8}
    G^{SC}(\omega) = \bigl(G^{-1}_{(0)}(\omega) - \tilde{\Sigma}_{S}(\omega) -  \tilde{\Sigma}_{I,l}(\omega) - \tilde{\Sigma}_{I,nl}^{SC}(\omega)\bigr)^{-1}.
\end{align} \\ 
Let us note that $\tilde{\Sigma}_{S}(\omega)$, originating from the system Hamiltonian, is always a purely real function and does not contribute to the dissipative dynamics of the system. 
Consequently, it does not play a role in the Fluctuation Dissipation relation, which is our next focus. Additionally, $\tilde{\Sigma}_{I,l}$, being independent of the system Green's function, always remains unchanged under the self-consistent Born approximation. Therefore, in proving the consistency of the Non-Crossing Approximation with the Fluctuation Dissipation Relation, the self-energy contribution $\tilde{\Sigma}_{I,nl}$ plays a key role. Hence, it is useful to rewrite this Dyson series as follows:
\begin{align} \label{lbl9}
        G^{SC}(\omega) = \bigl (G^{-1}_{1}(\omega)- \tilde{\Sigma}_{I,nl}^{SC}(\omega) \bigr)^{-1} ,
\end{align}
defining,
\begin{align} \label{lbl10}
    G_{1}(\omega) = \bigl (G^{-1}_{(0)}(\omega)- \tilde{\Sigma}_{S}(\omega) - \tilde{\Sigma}_{I,l}(\omega) \bigr )^{-1}.
\end{align}
Doing this allows us to view $G_{1}$ as a redefined bare Green's function, and write a Dyson series with respect to the non-linear interaction. Henceforth, unless stated otherwise, we refer to $G_{1}$ as our bare Green's function.

\subsection*{Consistency of Non-Crossing Approximation with the Fluctuation Dissipation Relation} \label{lblb4}
Here, focusing on the simple case for the non-linear 
interaction Hamiltonian as expressed by Eq.\eqref{lbl4}, we demonstrate that the self-consistent steady-state Green's function obeys the Fluctuation-Dissipation Relation (FDR), i.e.,
\begin{align}
    G^{SC}_{K}(\omega) = \bigl( G^{SC}_{R}(\omega) - G^{SC}_{A}(\omega) \bigr) \coth \Bigl(\frac{\beta}{2} \omega \Bigr) .
\end{align}
The arguments used to do so are proven in Appendix \ref{lblb10} and Appendix \ref{lblb11}. We have proved them for the general class of models described by Hamiltonians in Eq.\eqref{lbl1}. So, analogous proof demonstrates FDR for each of these models.

To prove this, we focus on an iterative mechanism of implementing the self-consistent Born approximation. Upon substituting the Dyson equation for $G^{SC}$ on the right-hand side in Eq.\eqref{lbl9}, the self-energy $\tilde{\Sigma}^{SC}_{I,nl}$ recursively appears on the right-hand side of the equation again. This is why, self-consistent Born approximation is termed ``self-consistent" \cite{altland2010condensed}. As a result, the self-consistent Born approximation can be implemented through the following iterative mechanism:

\textit{Initial Step:}  Write down the Dyson equation with the leading-order Green's function $G_{1}$ expressed by Eq.\eqref{lbl10}, appearing inside the leading order self-energy for the model under consideration, expressed by Eq.\eqref{lbl7}, as follows:
\begin{align} \label{lbl11}
    &\tilde{\Sigma}_{1,nl}(\omega) = \int_{\tau} D_{B}(\tau) \circ G_{1}(\tau) e^{ i \omega \tau} , \nonumber\\
    &G_{2}(\omega) = \bigl(G^{-1}_{1}(\omega) - \tilde{\Sigma}_{1,nl}^{(2)}(\omega)\bigr)^{-1} .
\end{align}
\textit{Iteration 1:} Define a new self-energy `$\tilde{\Sigma}_{2,nl}$' and a corresponding Green's function `$G_{3}$' with the Green's function $G_{2}$ appearing inside the self-energy:
\begin{align} \label{lbl12}
    &\tilde{\Sigma}_{2,nl}(\omega) = \int_{\tau} D_{B}(\tau) \circ G_{2}(\tau) e^{i \omega \tau} , \nonumber\\
    &G_{3}(\omega) = \bigl(G^{-1}_{1}(\omega) - \tilde{\Sigma}_{2,nl}((\omega)\bigr)^{-1}.
\end{align} 
\textit{Iteration 2:} Further, define a new self-energy `$\tilde{\Sigma}_{3,nl}$' and a corresponding Green's function `$G_{4}$' with the Green's function $G_{3}$ appearing inside the self-energy:
\begin{align} \label{lbl13}
    &\tilde{\Sigma}_{3,nl}(\omega) = \int_{\tau} D_{B}(\tau) \circ G_{3}(\tau) e^{i \omega \tau} \nonumber\\
    &G_{4}(\omega) = \bigl(G^{-1}_{1}(\omega) - \tilde{\Sigma}_{3,nl}(\omega)\bigr)^{-1}
\end{align}
\begin{center}. \end{center}
\begin{center} . \end{center}

\begin{figure}
\includegraphics[scale=.42]{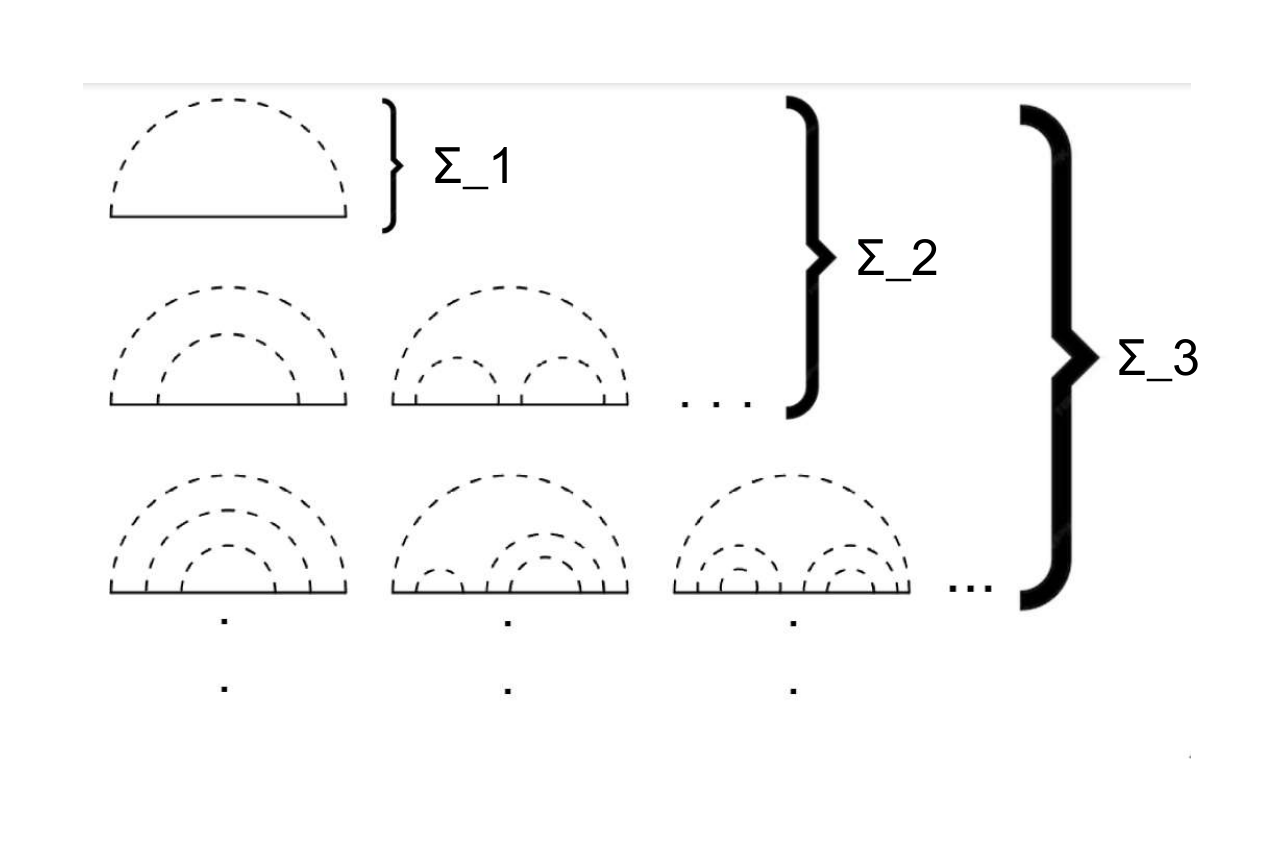}
	\caption{Non-crossing diagrams captured by the Self-Energy $\tilde{\Sigma}_{1,nl}$, $\tilde{\Sigma}_{2,nl}$ and $\tilde{\Sigma}_{3,nl}$, expressed by Eq.\eqref{lbl11}, Eq.\eqref{lbl12} and Eq.\eqref{lbl13} respectively. Here, the solid lines represent a system propagator, and the dashed lines represent a bath propagator.}\label{fig1}
\end{figure}
Performing this iteration infinitely generates all the NCA diagrams, and therefore, generates the self-consistent Green's function $G^{SC}$. This is a method for implementing the self-consistent Born approximation. Now, to proceed further and use this to show that $G^{SC}$ obeys the FDR, we have proved two key statements.
\begin{enumerate}
    \item \textit{Proved in Appendix \ref{lblb10}}: If the self-energy obeys the FDR, then the corresponding Green's function related to it by the Dyson series also obeys the same. \label{lb4}
    \item \textit{Proved in Appendix \ref{lblb11}}: If the Green's functions appearing inside the self-energy obey the FDR, then the self-energy itself also obeys the same. \label{lb3}
\end{enumerate}
So, for the case of non-linear interaction Hamiltonian as expressed by Eq.\eqref{lbl4}, by alternately applying Statement \ref{lb4} and Statement \ref{lb3}, one can observe that Green's function and self-energy generated at each step of the iterative process satisfy the FDR. Further, as depicted in FIG. \ref{fig1}, each subsequent iteration corresponds to a superclass of NCA diagrams with respect to the previous step. As mentioned earlier, performing this iteration infinitely generates all the NCA diagrams. Thus, this demonstrates that the self-consistent Green's function also satisfies the FDR. As previously stated, Statement \ref{lb4} and Statement \ref{lb3} have been proved abstractly for the class of models described by the Hamiltonian in Eq.\eqref{lbl1}, analogous proof demonstrates FDR for each of these models. 

\section{Comparison of greens function obtained using different techniques }\label{sec-IV}

\begin{figure}[h!]
	\centering
	\begin{subfigure}[b]{\linewidth}
        \includegraphics[width=\linewidth]{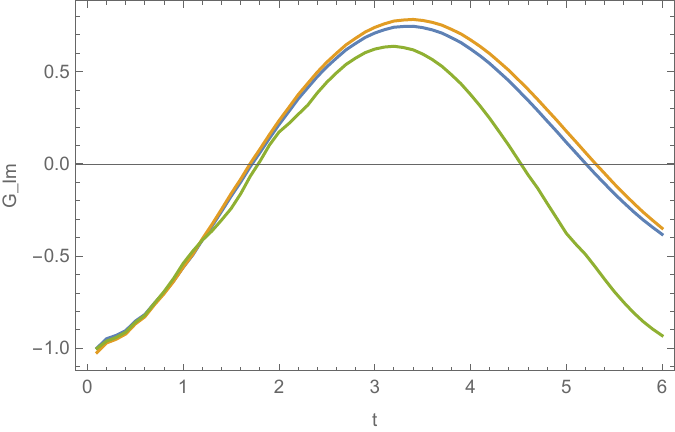}
		\caption{$ \Im(G_{R})$}
	\end{subfigure}\\
 \vfill
	\begin{subfigure}[b]{\linewidth}
		\includegraphics[width=\linewidth]{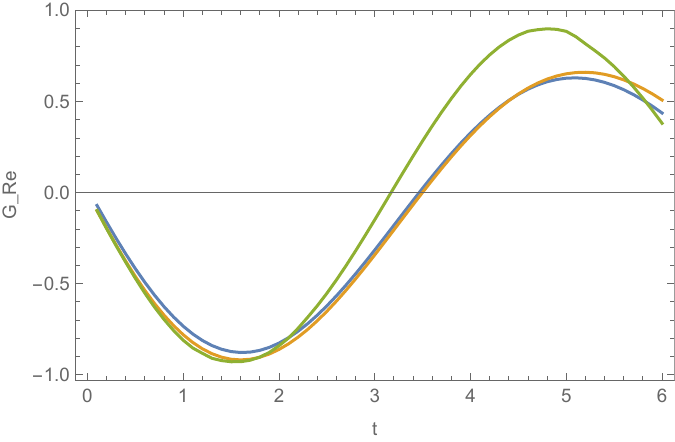}
		\caption{$\Re(G_{R}) $}
	\end{subfigure}
	\caption{This plot depicts comparison between the steady state Green's functions $G_{2,R}$, $G_{B,R}$, and $G^{SC}_{R}$ for $\lambda^{2} = 0.02$. The blue, orange and green lines represent $G_{B,R}$, $G_{2,R}$, and $G^{SC}_{R}$, respectively.}
	\label{fig3}
\end{figure}

\begin{figure}[h!]
	\centering
	\begin{subfigure}[b]{\linewidth}
        \includegraphics[width=\linewidth]{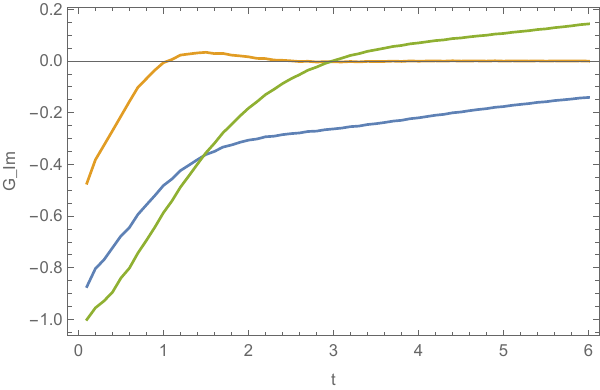}
		\caption{$ \Im(G_{R})$}
	\end{subfigure}\\
 \vfill
	\begin{subfigure}[b]{\linewidth}
		\includegraphics[width=\linewidth]{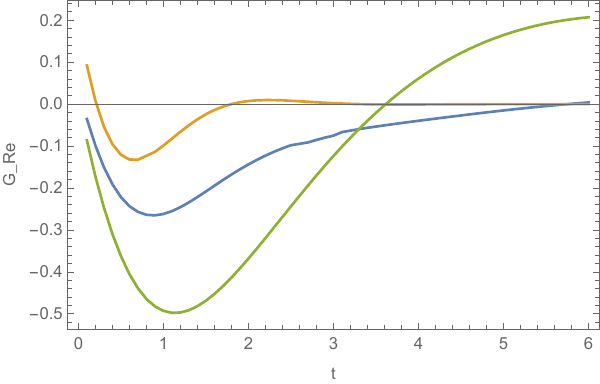}
		\caption{$\Re(G_{R}) $}
	\end{subfigure}
	\caption{This plot depicts comparison between $G_{B,R}$, $G_{2,R}$, and $G^{SC}_{R}$ for $\lambda^{2} = 0.2$. The blue, orange, and green lines represent $G_{B,R}$, $G_{2,R}$, and $G^{SC}_{R}$, respectively.}
	\label{fig4}
\end{figure}
In this section, we make a quantitative comparison between the self-consistent Green's function $G^{SC}$ given by Eq.\eqref{lbl9}, Green's function $G_{2}$ given by Eq.\eqref{lbl11}, and Green's function $G_{B}$ from the standard Born Approximation given by Eq.\eqref{lbl14}. The iterative mechanism for calculating the self-consistent Green's function discussed in the previous section highlights that Eq.\eqref{lbl9} is not of the first order but rather of infinite order. Hence, we emphasize that the self-consistent Green's function might remain accurate even when the Green's function from the `standard Born approximation' exhibits significant deviation. The self-consistent Green's function might also qualitatively capture the physics in the strong-coupling regime \cite{scarlatella2021self}, while quantitative accuracy cannot be expected there.

The Green's function $G^{SC}$ corresponds to the self-energy $\tilde{\Sigma}^{SC}$, which contains all the NCA diagrams. As depicted in FIG. \ref{fig1}, $G_{2}$ corresponds to the self-energy $\tilde{\Sigma}_{1}$, which represents the smallest class of NCA diagrams that satisfy the FDR. For the simple case of Hamiltonian given by Eq.\eqref{lbl4}, we explicitly demonstrate in Appendix \ref{lblb13} that $G_{2}$ satisfies the FDR. Conversely, in the same Appendix, we show that the steady-state Green's function $G_{B}$ from the `standard Born approximation' fails to satisfy the FDR.

Here, we perform a quantitative comparison of the steady-state Green's functions $G^{SC}(\tau)$, $G_{2}(\tau)$, and $G_{B}(\tau)$ for the Hamiltonian in Eq.\eqref{lbl4}. The explicit analytical derivation for $G_{2}(\tau)$ and $G_{B}(\tau)$ is provided in Appendix \ref{lblb13}. Meanwhile, we compute $G^{SC}(\tau)$ by numerically solving the integral-differential equation given by Eq.\eqref{lbl9}. Solving Eq.\eqref{lbl9} requires $G^{SC}(0)$ and $\frac{dG^{SC}}{d\tau}(\tau)|_{\tau=0}$ as initial conditions, which are steady state one-point expectation values. Determining the precise initial condition amounts to computing a density matrix, denoted as $\rho(t)$ using a self-consistent dynamical map (as detailed in paper \cite{scarlatella2021self}), and then taking the $t \rightarrow \infty$ limit. This has a high computational cost. So, instead of doing this, we use steady-state one-point expectation values associated with the Gibbs state,
\begin{align}
    \rho_{G} = \frac{e^{-\beta H_{S}}}{{\rm Tr} \bigl[e^{-\beta H_{S}}\bigr]} ,
\end{align}
to compute $G^{SC}(\tau)$. For weak coupling, we expect the initial condition $\rho_{G}$ should be very close to the correct one. This is further supported by the observation that $G^{SC}(\tau)$ closely agrees with the other two Green's functions in FIG. \ref{fig3}. In contrast, for strong coupling, we expect the correct initial condition to deviate from the one obtained from $\rho_{G}$. Therefore, by $G^{SC}(\tau))$ plotted in FIG. \ref{fig4} we aim to depict its qualitative behavior and quantitative precision isn't expected.

We consider here a thermal bath with the spectral density:
\begin{align}
    J(\omega) = \pi \frac{\omega}{1 + \omega^{2}} .
\end{align}

For comparison, we focus on the retarded Green's function of the system, given by:
\begin{equation}
       G_{R}(\omega) =  \bigl(G^{-1}_{1 R}(\omega)-\Tilde{\Sigma}_{nl,R}(\omega)\bigr)^{-1} ,
\end{equation}
FIG. \ref{fig3} demonstrates that for weaker coupling, with $\lambda^{2} = 0.02$, $G^{SC}(\tau)$, $G_{2}(\tau)$ and $G_{B}(\tau)$ closely agree with each other. On the other hand, in FIG. \ref{fig4} for stronger coupling, with $\lambda^{2} = 0.2$, these functions exhibit significant differences. As discussed earlier, this deviation is expected because each of the steady-state Green's functions, $G_{B}$, $G_{2}$, and $G^{SC}$, capture diagrams to different orders in perturbation theory. All these Green's functions display damped oscillatory behavior.

\section{Discussion }\label{sec-V}
In open quantum systems, thermalization implies that multi-time correlation functions must conform to the KMS/FDR condition. In this paper, we specifically explore the KMS condition for the two-point correlation function. In instances of exactly solvable models like Caldeira-Leggett, we can explicitly verify the satisfaction of the KMS condition by the two-time correlation function at late times. However, for non-exactly solvable models, approximate techniques are required to compute the two-point function.

The initial segment of our paper employs the image operator method as an approximate technique, illustrating that the resultant correlation function adheres to the KMS relation. Nonetheless, this method establishes KMS only for specific examples and is reliant on the weak system-bath coupling approximation. To go beyond the standard weak coupling limit, the subsequent section of our paper introduces a self-consistent perturbative technique—the self-consistent Born approximation or NCA approach—based on the Schwinger-Keldysh path integral. Through this approach, we abstractly show that the steady-state correlation function must conform to the KMS relation for a generic system. This approach also allows us to go beyond the standard weak coupling limit. To demonstrate this, we perform a quantitative comparison of the steady-state Green's functions from the NCA approach with the Green's function from the standard weak coupling limit. These functions closely agree for weak coupling, whereas differ significantly in the stronger coupling, demonstrating that the NCA approach can help to go beyond the weak coupling limit.

 It is worth mentioning that one can systematically go beyond the self-consistent Green's function approximation (NCA) by taking one crossing diagram (OCA). In other words, we can include a leading order diagram beyond the NCA to the self-energy and make a comparison similar to the one depicted in FIG. \ref{fig3} and FIG. \ref{fig4}.
 \begin{figure}
\includegraphics[scale=.3]{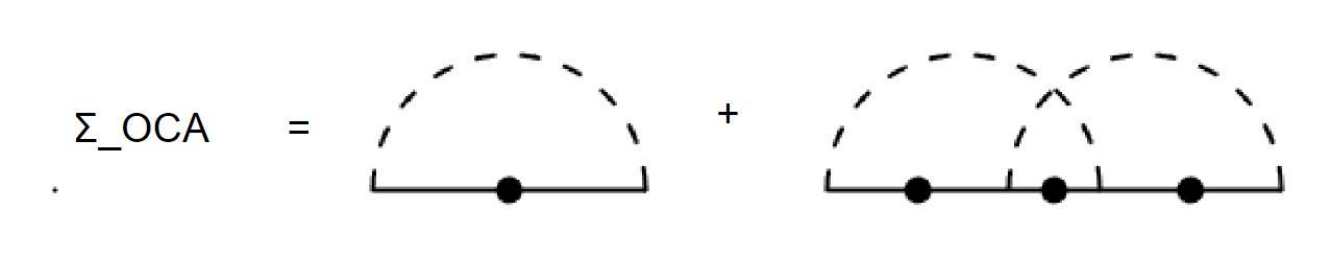}
	\caption{ Self Energy `$\tilde{\Sigma}_{OCA}$' for One Crossing Approcimation. Here, the solid line with a `dot' represents the total system Green's function, and the dashed line represents a bath Green's function. }
	\label{fig2}
\end{figure}
The self-energy corresponding to the OCA is depicted diagrammatically in FIG. \ref{fig2}. It will again be interesting to see whether this bigger class of diagrams satisfies the KMS condition or not. Further,  one can try to make a comparison between the accuracy of the steady state green's function derived using NCA, OCA, and higher-order self-consistent approximations \cite{scarlatella2021self}.  
The principle of Self-Consistency is not limited to expanding the two-point function. As an extension of our work, it will be interesting to check whether the self-consistent diagrams for four-point and higher-point functions satisfy the KMS condition. Along these lines, it would be interesting to investigate OTOC under the same set of ideas.

Another useful approximate way to calculate correlation function is the standard  Quantum Regression Theorem (QRT). The QRT asserts that understanding the time evolution of a single-point function is adequate for determining the time evolution of two-point or multi-point correlation functions. For calculating the one-point function and to go beyond the standard weak coupling limit, recently in \cite{scarlatella2021self}, the NCA-Master equation or  NCA Dynamical map was used. It is formally very similar and reduces to the standard master equations at sufficiently weak coupling.
An intriguing avenue of exploration would involve incorporating the NCA Dynamical maps with the QRT to compute Green's function. This dynamic analysis could then be compared with the steady-state Green's function calculated using the NCA approach outlined in this paper.

We expect that this conceptual framework, the NCA approach,  is extendable to out-of-equilibrium correlation functions \cite{agarwal2023initial}. Note that, to calculate the out-of-equilibrium correlation function we need information of the system's initial state. Additionally, it is known that out-of-equilibrium correlation functions break translational symmetry, rendering Fourier space analysis difficult. These two facts imply that, in general, calculating the out-of-equilibrium correlation function is technically challenging. However, we believe that this NCA technique is well-suited to go beyond the standard weak coupling limit in calculating the out-of-equilibrium correlation functions.

Note: All the authors have contributed equally to this work.

\section*{Acknowledgements} 
SK acknowledges the CSIR fellowship with Grant Number 09/0936(11643)/2021-EMR-I. LSR acknowledges the support of the INSPIRE fellowship from the Department of Science and Technology, Government of India. SK would like to thank Md. Ali and Katha Gangguli for useful discussions. LSR
would like to extend gratitude to Abhinav Dhawan and Sujal Kataria for the useful discussions.
We would also like to thank Bijay Kumar Agarwalla, Nishant Agarwal, and Yi-Zen Chu for extensive discussion.

\bibliographystyle{apsrev4-2}
 \bibliography{references.bib}
\appendix


\section{Image operator method}\label{appa}
In this appendix, we present a detailed derivation of the two-point correlation function using the image operator method.  It's important to note that our derivation applies to an arbitrary system Hamiltonian interacting with a bath through a generic system operator $S$. The total Hamiltonian for our configuration is given by the expression:
\begin{equation} \label{aeq1nma}
\begin{split}
H & = H_{S} + H_{R} + H_{S R} \\
& = H_{S} + \sum_{k} \Omega_{k} b_{k}^\dagger b_{k} + \sum_{k} \alpha_{k}( S b_{k}^\dagger + S^\dagger b_{k})\;,
\end{split}
\end{equation}
here $b_{k}$ and $b_{k}^\dagger$ represent bosonic or fermionic annihilation and creation operators, respectively. We want to calculate the two-point function of the form,
$ \langle O_{1}(t+\tau)O_{2}(t) \rangle$, at the steady state i.e. at $t \to \infty$.
To compute $ \langle O_{1}(t+\tau)O_{2}(t) \rangle$, we are going to first express it in terms of one-point reduced operators.
In Ref. \cite{karve2020heisenberg,Sakil1}, it was shown that the two-point reduced operator can be written  as
\begin{widetext}
\begin{equation}
\label{two-reduced}
	[O_{1  }(t+\tau)O_{2}(t)]_{S}=O_{1 S}(t+\tau)O_{2 S}(t)+ I[O_{1 S}(t+\tau),O_{2 S}(t)]\;.
\end{equation}
Here,  $O_{1 S} $ and $O_{2 S} $  represent the reduced one-point operators whose evolution is governed by the Born master equation \cite{breuer}. The term $I[O_{1 S}(t+\tau), O_{2 S}(t)]$ is referred to as the irreducible term, and it can be expressed in terms of the one-point reduced operator up to the leading order in system-bath coupling as 
\begin{align} \label{eqgI}
	I[O_{1 S}(t+\tau),O_{2 S }(t)]  \!&=\int_{0}^{t+\tau} \!d\tau_{1}\! \int_{0}^{t}\! d\tau_{2}\; \beta(\tau_{2}\!-\!\tau_{1}) \Big[ O_{1 S}(t+\tau),
	\Tilde{S}(-\tau_{1})\Big]\Big [ \Tilde{S}^\dagger(-\tau_{2}), O_{2 S}(t)
	\Big]\nonumber\\
	&+\int_{0}^{t+\tau} \!d\tau_{1}\! \int_{0}^{t}\! d\tau_{2}\; \alpha(\tau_{2}\!-\!\tau_{1})\Big[ O_{1 S}(t+\tau),
	\Tilde{S}^\dagger(-\tau_{1})\Big]\Big [ \Tilde{S}(-\tau_{2}), O_{2 S}(t)
	\Big]\;,
\end{align}
where $\Tilde{S}(t)$ represents the interaction picture operator, defined as $\Tilde{S}(t) = e^{-iH_{S}t} S e^{iH_{S}t}$. This operator can be decomposed as $\Tilde{S}(t) =\sum_{m} S_{m} e^{i\omega'_{m} t}$, whereas $\alpha(\tau)$ and $\beta(\tau)$ are intricately connected to the bath correlation function, or, to be more explicit,
\begin{align}
	\label{correlation1}
	&\alpha(\tau)=\sum_{k}  |\alpha_{k}|^2 \, {\rm Tr}_{R} \Big[b_{  k}\Tilde{b}^\dagger_{k}(-\tau)\rho_{R}\Big] =\sum_{k}  |\alpha_{k}|^2 \big(1-\eta \;n_{\eta}(\Omega_{k})\big)e^{i\Omega_{k}\tau},\nonumber\\
	& \beta(\tau)=\sum_{k}  |\alpha_{k}|^2 \, {\rm Tr}_{R} \Big[b^\dagger_{  k}\Tilde{b}_{k}(-\tau)\rho_{R}\Big] =\sum_{k}  |\alpha_{k}|^2  n_{\eta}(\Omega_{k}) e^{-i\Omega_{k}\tau}\;.
\end{align}
In this context, $n_{\eta}(\Omega)$ denotes the Bose or Fermi distribution functions, expressed as $n_{\eta}(\Omega)=[e^{\beta \Omega}+\eta]^{-1}$, where $\eta=+1$ corresponds to fermions, and $\eta=-1$ corresponds to bosons.
 Now, let's focus on the first term of the irreducible part, $I[O_{1 S}(t+\tau),O_{2 S}(t)]$, in Eq.\eqref{eqgI} i.e.
\begin{align}
   &\int_{0}^{t+\tau} \!d\tau_{1}\! \int_{0}^{t}\! d\tau_{2}\; \beta(\tau_{2}\!-\!\tau_{1}) \Big[ O_{1 S}(t+\tau),
	\Tilde{S}(-\tau_{1})\Big]\Big [ \Tilde{S}^\dagger(-\tau_{2}), O_{2 S}(t)
	\Big]\nonumber\\
 &=\sum_{j,l,m,n}\int_{0}^{t+\tau} \!d\tau_{1}\! \int_{0}^{t}\! d\tau_{2}\; \beta(\tau_{2}\!-\!\tau_{1}) \Big[ O_{1 j S}(t+\tau-\tau_{1}),
	S_{m}\Big]\Big [ S^\dagger_{n}, O_{2 l  S}(t-\tau_{2})
	\Big] e^{-i(\omega_{j}+\omega'_{m})\tau_{1}} e^{-i(\tilde{\omega}_{l}-\omega'_{n})\tau_{2}}+O(|\alpha_{k}|^4).
\end{align}
In the last step of the above equation, we insert the following expressions
\begin{align}
   & O_{1 S}(t)=\sum_{j}  O_{1 j  S}(t-t') e^{-i\omega_{j}t'}+O(|\alpha_{k}|^2),\nonumber\\
    & O_{2 S}(t)=\sum_{l}  O_{2 l  S}(t-t') e^{-i\tilde{\omega}_{l}t'}+O(|\alpha_{k}|^2)\;.
\end{align}
Let's define, $\tau'_{1}=t+\tau-\tau_{1}$ and $\tau'_{2}=t-\tau_{2}$,
 from the definition of $\tau'_{1} $ it is clear that the limit of $ \tau'_{1}$ will be from $t+\tau$ to $0$ and similarly  the limit of $ \tau'_{2}$ will be from $t$ to $0$. In terms of these new variables, we get the following expression
\begin{align}
   &\sum_{j,l,m,n}\int_{0}^{t+\tau} \!d\tau_{1}\! \int_{0}^{t}\! d\tau_{2}\; \beta(\tau_{2}\!-\!\tau_{1}) \Big[ O_{1 j S}(t+\tau-\tau_{1}),
	S_{m}\Big]\Big [ S^\dagger_{n}, O_{2 l  S}(t-\tau_{2})
	\Big] e^{-i(\omega_{j}+\omega'_{m})\tau_{1}} e^{-i(\tilde{\omega}_{l}-\omega'_{n})\tau_{2}}\nonumber\\
&=\sum_{j,l,m,n}\int_{0}^{t+\tau} \!d\tau'_{1}\! \int_{0}^{t}d\tau'_{2}\;\beta(\tau'_{1}\!-\!\tau'_{2}-\tau) \Big[ O_{1 j S}(\tau'_{1}),
	S_{m}\Big]\Big [ S^\dagger_{n}, O_{2 l  S}(\tau'_{2})
	\Big] e^{-i(\omega_{j}+\omega'_{m})(t+\tau-\tau'_{1})} e^{-i(\tilde{\omega}_{l}-\omega'_{n})(t-\tau'_{2})}\;.
\end{align}
Similarly, we can simplify the other term of the irreducible part $I[O_{1 S}(t+\tau), O_{2 S}(t)]$, expressed in Eq.\eqref{eqgI}. Finally, by substituting $\alpha$, $\beta$ defined in Eq.\eqref{correlation1}, we get the two-point reduced operator formulated in Eq.\eqref{eq2ptst}.
\section{Laplace transformation}\label{appb}
In this appendix, we are going to find the Laplace transformation of $ a_{S}(t)$. To do that we are going to first write down the equation of motion of $ a_{S}(t)$
\begin{align}
    \frac{d}{dt}a_{S}(t)=-i\omega_{0}a_{S}(t)-\int^{t}_{0} dt' K(t-t')a_{S}(t')\;,
\end{align}
where $K(t-t')=\sum_{k}|\alpha_{k}|^2 e^{-i\omega_{k}(t-t')}$. Laplace transformation of the above equation gives
\begin{align}
    \tilde{a}_{S}(-i\Omega)=\frac{a}{i(\omega_{0}-\Omega)+\tilde{K}(-i\Omega)}\;,
\end{align}
where $\tilde{K}(-i\Omega)=J(\Omega)/2+i \Sigma(\Omega)$.
\section{Schwinger Keldysh Path Integral} \label{appc}
The total Hamiltonian of Eq.\eqref{lbl1} when promoted to the Schwinger-Keldysh Path Integral and expressed in the classical-quantum basis for field operators takes the form $S = S_{S} + S_{B} +S_{I,l} + S_{I,nl}$, where
\begin{align}\label{lbl16}
    &S_{I,l} = \int_{\tau} \bigl( a^{*}_{c}(\tau)b_{k,q}(\tau) + a^{*}_{q}(\tau)b_{k,c}(\tau)  + h.c.  \bigr) ,\nonumber \\
    &S_{I,nl} = \int_{\tau} \bigl( (a^{*m}a^{n})_{c}(\tau)b_{k,q}(\tau)+(a^{* m}a^{n})_{q}(\tau)b_{k,c}(\tau)  + h.c. \bigr).
\end{align}
Here, $a$ and $b$ are complex-valued fields associated with the coherent states of the system and bath, respectively. As the Schwinger-Keldysh functional integral, with a total action involving both system and bath degrees of freedom, is quadratic in the latter, the bath can be integrated out. After integrating out the bath, the reduced Schwinger-Keldysh action is,  $S=S_{s}+S'_{l}+S'_{nl}$, where
\begin{align} \label{lbl17}
    &    S'_{l} = \int_{w} 
                 \begin{pmatrix}
                    a_{c}^{*}(\omega) & a_{q}^{*}(\omega)
                 \end{pmatrix}
                 \begin{pmatrix}
                   -2i J(\omega) (2 n(\omega) + 1 ) & -i J(\omega) + \Sigma(\omega)
                   \\
                   i J(\omega) + \Sigma(\omega) & 0
                 \end{pmatrix}                 
                \begin{pmatrix}
                    a_{c}(\omega) \\
                    a_{q}(\omega)
                \end{pmatrix} ,
                 \nonumber\\
    &    S'_{nl} = \int_{w} 
                 \begin{pmatrix}
                (a^{*m}a^{n})_{c}(\omega) (a^{*m}a^{n})_{q}(\omega)
                 \end{pmatrix} \begin{pmatrix}
                   -2i J(\omega) (2 n(\omega) + 1 ) & -i J(\omega) + \Sigma(\omega)
                   \\
                   i J(\omega) + \Sigma(\omega) & 0
                 \end{pmatrix}                 
                \begin{pmatrix}
                    (a^{*n}a^{m})_{c}(\omega) \\
                    (a^{*n}a^{m})_{q}(\omega)
                \end{pmatrix} .
\end{align}
\end{widetext}
 
\section{FDR : Self-Energy $\leftrightarrow$ Green's Function} \label{lblb10}
In this section, we will demonstrate that if the self-energy $\tilde{\Sigma}_{I}$ due to system bath interaction satisfies the Fluctuation-Dissipation Relation (FDR), i.e., 
\begin{align}
    \tilde{\Sigma}_{K}(\omega) = \bigl(\tilde{\Sigma}_{R}(\omega) - \tilde{\Sigma}_{A}(\omega) \bigr) \coth \Bigl(\frac{\beta}{2} \omega \Bigr) ,
\end{align}
then the corresponding Green's function $G$, related to it by the Dyson series:
\begin{align}
    G(\omega) = \bigl(G_{(0)}^{-1}(\omega) - \tilde{\Sigma}_{I}(\omega) \bigr)^{-1},
\end{align}
must also obey the same relation. Here, $G_{(0)}$ is the bare Green's function with respect to the system-bath-interaction, i.e., calculated using $H_{S}$, and it is given by
\begin{align}
    G_{(0)}(\omega) = 
    \begin{pmatrix}
        0 & \frac{1}{\omega - \Sigma_{S}(\omega)} \\ \frac{1}{\omega - \Sigma_{S}(\omega)} & 0
    \end{pmatrix},
\end{align}
where `$\Sigma_{S}$' is the self-energy due to intra-system interactions, making it always a real-valued function. Using the Dyson equation, it is easy to show that
\begin{align}
       &G(\omega) =  \bigl(G^{-1}_{(0)}(\omega)-\tilde{\Sigma}_{I}(\omega)\bigr)^{-1}\nonumber \\ &=\begin{pmatrix}
                  \frac{   \tilde{\Sigma}_{IK}(\omega)}{( \omega- \Sigma_{S}(\omega) - \tilde{\Sigma}_{IA}(\omega)) ( \omega - \Sigma_{S}(\omega) -\tilde{\Sigma}_{IR}(\omega))}  & \frac{1}{\omega - \Sigma_{S}(\omega)-\tilde{\Sigma}_{IR}(\omega)} 
                   \\
                    \frac{1}{\omega - \Sigma_{S}(\omega)-\tilde{\Sigma}_{IA}(\omega)} & 0
                 \end{pmatrix} .
\end{align}
That implies,
\begin{align} \label{lbl99}
    \frac{G_{K}(\omega)}{G_{R}(\omega) - G_{A}(\omega)} =  \frac{\tilde{\Sigma}_{I K}(\omega)}{\tilde{\Sigma}_{I R}(\omega) - \tilde{\Sigma}_{I A}(\omega)} .
\end{align}
From here, it's transparent that Green's function $G$ satisfies FDR if and only if $\tilde{\Sigma}_{I}$ satisfies the same.

\section{Proof of FDR for the general form of leading order self-energy} \label{lblb11}
\begin{figure}[h!]
	\centering
	\begin{subfigure}[b]{\linewidth}
        \includegraphics[width=4 cm]{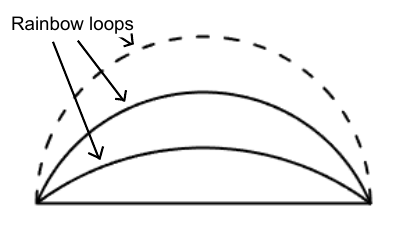}
		\caption{Diagrams with no Rings}
	\end{subfigure}\\
 \vfill
	\begin{subfigure}[b]{\linewidth}
		\includegraphics[width= 6cm]{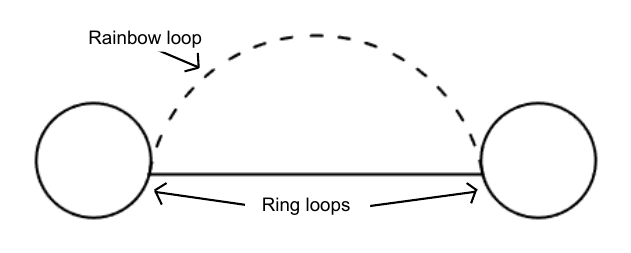}
		\caption{Diagrams with Rings}
	\end{subfigure}
	\caption{The two classes of Feynman diagrams for the simple case of the non-linear interaction Hamiltonian $H_{I, nl} = a^{\dagger 2} a b_{k} + a^{\dagger} a^{2} b_{k}^{\dagger}$. We define `rainbow' as a loop that corresponds to two distinct vertices, whereas, we define `ring' as a loop that has only one vertex associated with it. Here, the solid lines represent the system Green's functions, and the dashed lines represent the bath Green's functions.}
	\label{fig5}
\end{figure}

In this section, working with the general class of models represented by the Action in Eq. \eqref{lbl17}, we demonstrate a condition under which the leading-order self-energy term satisfies the FDR. To do that, we first identify that the loops appearing in the leading-order self-energy diagrams can be classified into `rainbow' and `rings'. We can classify all the Feynman diagrams into diagrams with a ring, and diagrams without a ring. This classification is illustrated in FIG. \ref{fig5} for the total Hamiltonian of the form given by Eq. \eqref{lbl1}, with the simple non-linear interaction Hamiltonian $H_{I, nl} = a^{\dagger 2} a b_{k} + a^{\dagger} a^{2} b_{k}^{\dagger}$. 

It will turn out, that only the rainbow part of these diagrams plays a non-trivial role in consistency with the FDR. The contribution from rings simply factors out from all the various time-ordered self-energies and, therefore, is inconsequential for the consistency with the FDR. Consequently, for simplicity, we first focus on the diagrams with no rings in the first subsection and prove the FDR. We then turn our attention to diagrams with rings in the next subsection.

\subsubsection{Diagrams with no rings}

Using the Schwinger-Keldysh path integral and employing the standard tool of Feynman diagrams that arise from the conventional Wick contraction, one can demonstrate that the leading-order self-energy diagrams with no rings (so only rainbow loops) are of the following general form:
\begin{align}
    \tilde{\Sigma}(\tau) = G_{1}(\tau_{1}) \circ G_{2}( \tau_{2}) \circ ... \circ G_{l}(\tau_{l}),
\end{align}
where $\tau_{j} \in \{ \tau, -\tau \}$ for all $j$, and $G_{j}(\tau_{j})$ represents a system or bath Green's function. We establish the rule for the product `$\circ$' in accordance with the rules for Wick contraction. The product `$\circ$' is defined as follows: \\
\begin{widetext}
for $ L = \{ 1,2,...,l \} $ ,
\begin{align}
    \tilde{\Sigma}_{K}(\tau) \! = \! -i \! \! \! \sum_{S \subseteq L \atop n(S) \epsilon 2 \mathbb{Z}} \! \! \biggl( \Bigl(\prod_{\mu \epsilon S} i &G_{K \mu}(\tau_{\mu}) \Bigr) \Bigl( \! \! \prod_{\nu \epsilon \overline{S} \atop \tau_{\nu} = \tau} \! \! i G_{R \nu}(\tau) \Bigr) \Bigl(\! \! \prod_{\nu \epsilon \overline{S} \atop \tau_{\nu} = -\tau} \! \! \!i G_{A \nu}(-\tau) \Bigr) \biggr)  - i \! \! \! \sum_{S \subseteq L \atop n(S) \epsilon 2 \mathbb{Z}} \! \! \biggl ( \Bigl(  \prod_{\mu \epsilon S}   i G_{K \mu}(\tau_{\mu}) \Bigr) \Bigl( \! \! \prod_{\nu \epsilon \overline{S} \atop \tau_{\nu} = \tau} \! \! i G_{A \nu}(\tau) \Bigr) \Bigl(\! \! \prod_{\nu \epsilon \overline{S} \atop \tau_{\nu} = -\tau} \! \! i G_{R \nu}(-\tau) \Bigr) \biggr) , \nonumber \\
    &\tilde{\Sigma}_{R}(\tau) = -i \! \! \! \! \!  \sum_{S \subseteq L \atop n(S) \epsilon 2 \mathbb{Z} + 1} \! \! \! \! \! \biggl( \Bigl( \prod_{\mu \epsilon S} i G_{K \mu}(\tau_{\mu}) \Bigr) \Bigl( \! \! \prod_{\nu \epsilon \overline{S} \atop \tau_{\nu} = \tau} \! \! i G_{R \nu}(\tau) \Bigr) \Bigl( \! \! \prod_{\nu \epsilon \overline{S} \atop \tau_{\nu} = -\tau}\! \! i G_{A \nu}(-\tau) \Bigr) \biggr) , \nonumber \\
    &\tilde{\Sigma}_{A}(\tau) = -i \! \! \! \! \!  \sum_{S \subseteq L \atop n(S) \epsilon 2 \mathbb{Z} + 1} \! \! \! \! \! \biggl( \Bigl( \prod_{\mu \epsilon S} i G_{K \mu}(\tau_{\mu}) \Bigr) \Bigl( \! \! \prod_{\nu \epsilon \overline{S} \atop \tau_{\nu} = \tau} \! \! i G_{A \nu}(\tau) \Bigr) \Bigl( \! \! \prod_{\nu \epsilon \overline{S} \atop \tau_{\nu} = -\tau}\! \! i G_{R \nu}(-\tau) \Bigr) \biggr) , 
\end{align}   
\end{widetext}
where, $n(S)$ denotes the number of elements in the set $S$, $\overline{S}$ denotes the set complement to $S$, and $2 \mathbb{Z}$ and $2 \mathbb{Z} + 1$ denote the sets of all even and odd integers, respectively. We represent the self-energy matrix in terms of the time-ordered self-energies as follows:
\begin{align}
    \tilde{\Sigma}(\tau) =
    \begin{pmatrix}
         0 &  \tilde{\Sigma}_{A}(\tau) \\ \tilde{\Sigma}_{R}(\tau) & \tilde{\Sigma}_{K}(\tau)
    \end{pmatrix} .
\end{align}
As described in the paper \cite{chaudhuri2019spectral}, the contour-ordered thermal two-point correlators are related by the following structure:
\begin{align}
    \begin{pmatrix}
        G_{K}(\tau) & G_{R}(\tau) \\ G_{A}(\tau) & 0
    \end{pmatrix} =
        \int_{\omega} \rho(\omega)
        \begin{pmatrix}
            \coth(\frac{\beta}{2}\omega) & \theta(\tau) \\ -\theta(-\tau) & 0
    \end{pmatrix} e^{- i \omega \tau} .
\end{align}
Where, `$\rho(\omega)$' stands for `spectral function' which is directly related to the Fourier Transform of commutators in the theory:
\begin{align}
    \int_{\omega}\rho(\omega) e^{- i \omega \tau}= \langle [\ \!   \hat{a}^{\dagger}(\tau), \hat{a}(0)
     ]\ \! \! \rangle .
\end{align}
So, for $j$ such that $\tau_{j} = \tau$, 
\begin{align}
    G_{j}(\tau) &= 
    \begin{pmatrix}
        G_{K j}(\tau) & G_{R j}(\tau) \\ G_{A j}(\tau) & 0
    \end{pmatrix} \nonumber \\
    &= \int_{\omega_{j}} \! \! \! \rho_{j}(\omega_{j})
    \begin{pmatrix}
        \coth(\frac{\beta}{2}\omega_{j}) & \theta(\tau) \\ -\theta(-\tau) & 0
    \end{pmatrix} e^{- i \omega_{j} \tau} .
\end{align}
Next, for the convenience of denoting the product `$\circ$', we perform the following transformation for $G_{j}(\tau_{j})$ for which, $\tau_{j} = -\tau$:
\begin{align}
    &G_{j}(-\tau) = 
    \begin{pmatrix}
        G_{K j}(-\tau) & G_{A j}(-\tau) \\ G_{R j}(-\tau) & 0
    \end{pmatrix} \nonumber \\
    &= \int_{-\omega_{j}} \Bigl( -\rho_{j}\bigl(-(-\omega_{j})\bigr) \Bigr) 
    \begin{pmatrix}
        \coth (\frac{\beta}{2}\bigl(-\omega_{j})\bigr) & \theta(\tau) \\ -\theta(-\tau) & 0
    \end{pmatrix} e^{- i (-\omega_{j}) \tau}  \nonumber \\
    &= \int_{\omega_{j}'} \rho'_{j}(\omega'_{j}) 
    \begin{pmatrix}
        \coth(\frac{\beta}{2}\omega_{j}') & \theta(\tau) \\ -\theta(-\tau) & 0
    \end{pmatrix} e^{- i \omega_{j}' \tau} ,
\end{align}
where we define $\omega_{j}' = -\omega_{j}$ and $\rho_{j}'(\omega) = -\rho_{j}(-\omega)$. It is important to note that it assumes the same form as $G_{j}(\tau)$. This equivalence enables us to consistently express the self-energy term with no rings, as follows:
\begin{widetext}
\begin{align}
    \tilde{\Sigma}(\tau) = G_{1}(\tau) \circ G_{2}(\tau) \circ ... \circ G_{l}(\tau)
\end{align}
which gives, for $ L = \{ 1,2,...,l \} $ ,
\begin{align}
    \tilde{\Sigma}_{K}(\tau) \! = \! -i \! \! \! \sum_{S \subseteq L \atop n(S) \epsilon 2 \mathbb{Z}} \! \! \biggl( \Bigl(\prod_{\mu \epsilon S} i G_{K \mu}(\tau) \Bigr) \Bigl(  \prod_{\nu \epsilon \overline{S}}  i G_{R \nu}(\tau) \Bigr) \biggr)  - i \! \! \! \sum_{S \subseteq L \atop n(S) \epsilon 2 \mathbb{Z}} \! \! \biggl ( \Bigl(  \prod_{\mu \epsilon S}   i G_{K \mu}(\tau) \Bigr) \Bigl(  \prod_{\nu \epsilon \overline{S}}  i G_{A \nu}(\tau) \Bigr)\biggr) , 
\end{align}   
\begin{align}
    \tilde{\Sigma}_{R}(\tau) = -i \! \! \! \! \!  \sum_{S \subseteq L \atop n(S) \epsilon 2 \mathbb{Z} + 1} \! \! \! \! \! \biggl( \Bigl( \prod_{\mu \epsilon S} i G_{K \mu}(\tau) \Bigr) \Bigl(  \prod_{\nu \epsilon \overline{S}}  i G_{R \nu}(\tau) \Bigr) \biggr) ,
\end{align}
\begin{align}
    \tilde{\Sigma}_{A}(\tau) = -i \! \! \! \! \!  \sum_{S \subseteq L \atop n(S) \epsilon 2 \mathbb{Z} + 1} \! \! \! \! \! \biggl( \Bigl( \prod_{\mu \epsilon S} i G_{K \mu}(\tau) \Bigr) \Bigl(  \prod_{\nu \epsilon \overline{S}} i G_{A \nu}(\tau) \Bigr) \biggr)
\end{align}
Here, expressing the thermal Green's functions in the spectral function representation, one obtains,
\begin{align}
    \tilde{\Sigma}_{K}(\tau) = \int_{\omega_{1},\omega_{2},...,\omega_{l}}
    \Bigl (\prod_{j \epsilon L} \rho_{j}(\omega_{j}) \Bigr) \Bigl( \sum_{S \subseteq L \atop n(S) \epsilon 2 \mathbb{Z}} \bigl(\prod_{j' \epsilon \overline{S}} \coth(\frac{\beta}{2} \omega_{j'})\bigr ) \Bigr) e^{-i (\sum_{j \epsilon L} \omega_{j} \tau)}, 
\end{align}
\begin{align}
    \tilde{\Sigma}_{R}(\tau) = \int_{\omega_{1},\omega_{2},...,\omega_{l}} \theta(\tau) \Bigl(\prod_{j \epsilon L} \rho_{j}(\omega_{j})\Bigr) \Bigl( \sum_{S \subseteq L \atop n(S) \epsilon 2 \mathbb{Z} + 1 } \bigl(\prod_{j' \epsilon \overline{S}} \coth(\frac{\beta}{2} \omega_{j'})\bigr) \Bigr) e^{-i (\sum_{j \epsilon L} \omega_{j} \tau)},  
\end{align}
\begin{align}
    \tilde{\Sigma}_{A}(\tau) = \int_{\omega_{1},\omega_{2},...,\omega_{l}} - \theta(-\tau) \Bigl(\prod_{j \epsilon L} \rho_{j}(\omega_{j}) \Bigr) \Bigl( \sum_{S \subseteq L \atop n(S)\epsilon 2 \mathbb{Z} + 1} \bigl(\prod_{j' \epsilon \overline{S}} \coth(\frac{\beta}{2} \omega_{j'})\bigr )\Bigr) e^{-i (\sum_{j \epsilon L} \omega_{j} \tau)}. 
\end{align}
The self-energy adheres to the spectral function representation, as can be observed by employing the identity
\begin{align}
    \coth \bigl(\sum_{i = 1}^{l} x_{i} \bigr) = \frac{\sum_{S \subseteq L \atop  n(S) \epsilon 2 \mathbb{Z}} \{ \prod_{i \epsilon \overline{S}} \coth(x_{i}) \}}{\sum_{S \subseteq L \atop  n(S) \epsilon 2 \mathbb{Z} + 1} \{ \prod_{i' \epsilon \overline{S}} \coth(x_{i'}) \}} ,
\end{align}
and defining its spectral function as,
\begin{align}
    \varrho(\omega) = \int_{ \omega_{1}, \omega_{2}, ... ,\omega_{l}}
    \bigl(\prod_{j \epsilon L} \rho_{j}(\omega_{j}) \bigr) \delta \bigl(\omega - \sum_{j \epsilon L} \omega_{j} \bigr) ,
\end{align}
giving,
\begin{align}
    \begin{pmatrix}
         0 & \tilde{\Sigma}_{A}(\tau) \\  \tilde{\Sigma}_{R}(\tau) & \tilde{\Sigma}_{K}(\tau)
    \end{pmatrix} = \int_{\omega} \varrho(\omega)
    \begin{pmatrix}
         0 &  \theta(\tau) \\ -\theta(-\tau) & \coth(\frac{\beta}{2}\omega)
    \end{pmatrix} e^{- i \omega \tau} .
\end{align}
This proves the FDR for leading-order self-energy rainbow diagrams when the Green's functions appearing in the self-energy obey the same.

\subsubsection{Diagrams with Rings} \label{subsecb} 
Now, let's consider leading-order self-energy diagrams that also include rings. Firstly, due to the hermiticity of the interaction Hamiltonian, for any diagram with a ring involving the advanced Green's function, there exists a corresponding diagram with the ring involving the retarded Green's function. Therefore, the net contribution of these diagrams is proportional to $G_{R}(0)+G_{A}(0)$. Since $G_{R}(0)+G_{A}(0)=0$ \cite{kamenev2023field}, these diagrams cancel out, leaving only the diagrams with all the rings involving Keldysh Green's function. Interestingly, these self-energy diagrams have the same general form as sunset diagrams with additional factors of the ring's Keldysh Green's functions, i.e., $G_{mK}(0)$. So, for a self-energy diagram with `n' rings:
\begin{align}
    \tilde{\Sigma}(\tau) = \Bigl(\prod^{n}_{m = 1 }G_{m K}(0) \Bigr) G_{1}(\tau_{1}) \circ G_{2}(\tau_{2}) \circ ... \circ G_{l}(\tau_{l}), 
\end{align}
where $\tau_{j} \epsilon \{ \tau, -\tau \}$ for all $j$. Because the ring Green's functions appear as common factors in all the time-ordered self-energies, the previous section's proof for FDR follows and we get:
\begin{align}
    \varrho(\omega) = \Bigl(\prod^{n}_{m = 1 }G_{m K}(0) \Bigr) \int_{ \omega_{1}, \omega_{2}, ... ,\omega_{l}}
    \bigl(\prod_{j \epsilon L} \rho_{j}(\omega_{j}) \bigr) \delta \bigl(\omega - \sum_{j \epsilon L} \omega_{j} \bigr)
\end{align}
giving,
\begin{align}
    \begin{pmatrix}
         0 & \tilde{\Sigma}_{A}(\tau) \\  \tilde{\Sigma}_{R}(\tau) & \tilde{\Sigma}_{K}(\tau)
    \end{pmatrix} = \int_{\omega} \varrho(\omega)
    \begin{pmatrix}
         0 &  \theta(\tau) \\ -\theta(-\tau) & \coth(\frac{\beta}{2}\omega)
    \end{pmatrix} e^{- i \omega \tau} .
\end{align}
This completes the proof that if the Green's functions that appear inside the leading-order self-energy obey the FDR, then the self-energy obeys the same.
\section{Analytical calculation of steady state Green's functions $G_{B}$ and $G_{2}$} \label{lblb13}

In this section, we derive the explicit form of the steady-state Green's functions $G_{B}$ and $G_{2}$, expressed through the Dyson equations in Eq. \eqref{lbl14} and \eqref{lbl11}, respectively. Additionally, we demonstrate that $G_{2}$ satisfies the Fluctuation-Dissipation Relation (FDR), whereas $G_{B}$ fails to satisfy it.

Using the rules for product `$\circ$' as stated in Appendix \ref{lblb11},
the $\tilde{\Sigma}_{B}$ (Born) and $\tilde{\Sigma}_{1,nl}^{(2)}$ self energies defined in Section IV are:
\begin{align}
    &\tilde{\Sigma}_{B K}(\tau) = i (D_{B K}(\tau) G_{(0)K}(\tau) + D_{B R}(\tau) G_{(0)R}(\tau) + D_{B A}(\tau) G_{(0)A}(\tau)) \nonumber \\  &= - i \sum_{k} |\alpha_{k}|^{2} e^{-i (\omega_{0} + \omega_{k}) \tau} \xlongrightarrow{\text{F.T.}} - 2 i J(\omega - \omega_{0})  \nonumber \\
    &\tilde{\Sigma}_{B R}(\tau) = i (D_{B K}(\tau) G_{(0)R}(\tau) + D_{B R}(\tau) G_{(0)K}(\tau))  \nonumber \\  &= -i \theta(\tau) \sum_{k} |\alpha_{k}|^{2} \coth(\frac{\beta}{2}\omega_{k}) e^{- i (\omega_{0} + \omega_{k}) \tau} \xlongrightarrow{\text{F.T.}} -i J(\omega - \omega_{0}) \coth(\frac{\beta}{2}(\omega- \omega_{0} )) + \mathcal{P}\int_{\omega'} \frac{J(\omega - \omega_{0}) \coth(\frac{\beta}{2}(\omega'- \omega_{0} ))}{ \pi (\omega - \omega')}  \nonumber \\
    &\tilde{\Sigma}_{B A}(\tau) = i (D_{B K}(\tau) G_{(0)A}(\tau) + D_{B A}(\tau) G_{(0)K}(\tau))  \nonumber \\  &= i \theta(-\tau) \sum_{k} |\alpha_{k}|^{2} \coth(\frac{\beta}{2}\omega_{k}) e^{- i (\omega_{0} + \omega_{k}) \tau}  \xlongrightarrow{\text{F.T.}} i J(\omega - \omega_{0}) \coth(\frac{\beta}{2}(\omega- \omega_{0} )) + \mathcal{P}\int_{\omega'} \frac{J(\omega - \omega_{0}) \coth(\frac{\beta}{2}(\omega'- \omega_{0} ))}{ \pi (\omega - \omega')} 
\end{align}
Clearly,
\begin{align}
    \tilde{\Sigma}_{B K}(\omega) = (\tilde{\Sigma}_{B R}(\omega) - \tilde{\Sigma}_{B A}(\omega))\frac{1}{\coth(\frac{\beta}{2}(\omega- \omega_{0}))}
\end{align}
Note that, it does not satisfy the FDR. As a result, as proved in Appendix \ref{lblb10}, the corresponding Green's function $G_{B}$ also doesn't satisfy the FDR.
Similarly, if 
\begin{align}
    G_{1 K}(\tau) = \int_{\omega} \rho(\omega)
    \begin{pmatrix}
         \coth(\frac{\beta}{2}\omega) & \theta(\tau)  \\  -\theta(-\tau)  & 0
    \end{pmatrix} e^{-i\omega \tau}
\end{align}

Then, 
\begin{align}
    &\tilde{\Sigma}_{1 K}^{(2)}(\tau) = i (D_{B K}(\tau) G_{1K}(\tau) + D_{B R}(\tau) G_{1R}(\tau) + D_{B A}(\tau) G_{1A}(\tau)) \nonumber \\ 
     &= \int_{\omega}  \rho(\omega) \sum_{k} |\alpha_{k}|^{2} (\coth(\frac{\beta}{2}\omega_{k})\coth(\frac{\beta}{2}\omega)+1)e^{-i (\omega+\omega_{k})\tau}\nonumber \\ &\xlongrightarrow{\text{F.T.}} -2i \sum_{i} J(\omega-\omega_{i})(\coth(\frac{\beta}{2}(\omega - \omega_{i}))\coth(\frac{\beta}{2}\omega_{i}) + 1)  \nonumber \\    
    &\tilde{\Sigma}_{1 R}^{(2)}(\tau) = i (D_{B K}(\tau) G_{1R}(\tau) + D_{B R}(\tau) G_{1 K}(\tau)) \nonumber \\ &= -i \theta(\tau) \sum_{i} r_{i} \sum_{k} |\alpha_{k}|^{2} (\coth(\frac{\beta}{2}\omega_{k}) + \coth(\frac{\beta}{2}\omega_{i}) )e^{- i (\omega_{i} + \omega_{k}) \tau} \nonumber \\
    &\xlongrightarrow{\text{F.T.}}  \sum_{i} (-i J(\omega - \omega_{i})(\coth(\frac{\beta}{2}(\omega-\omega_{i}))+\coth(\frac{\beta}{2}\omega_{i})) + \mathcal{P} \int_{\omega'}  \frac{J(\omega' - \omega_{i})(\coth(\frac{\beta}{2}(\omega'-\omega_{i}))+\coth(\frac{\beta}{2}\omega_{i}))}{\pi (\omega - \omega')})\nonumber \\
    &\tilde{\Sigma}_{1 A}^{(2)}(\tau) = i (D_{B K}(\tau) G_{1A}(\tau) + D_{B A}(\tau) G_{1 K}(\tau))  \nonumber \\ &= i \theta(-\tau) \sum_{i} r_{i}^{*} \sum_{k} |\alpha_{k}|^{2} (\coth(\frac{\beta}{2}\omega_{k}) + \coth(\frac{\beta}{2}\omega_{i}^{*}) )e^{- i (\omega_{i}^{*} + \omega_{k}) \tau} \nonumber \\
    &\xlongrightarrow{\text{F.T.}}  \sum_{i} (-i J(\omega - \omega_{i})(\coth(\frac{\beta}{2}(\omega-\omega_{i}))+\coth(\frac{\beta}{2}\omega_{i})) + \mathcal{P} \int_{\omega'}  \frac{J(\omega' - \omega_{i})(\coth(\frac{\beta}{2}(\omega'-\omega_{i}))+\coth(\frac{\beta}{2}\omega_{i}))}{\pi (\omega - \omega')})\nonumber     
\end{align}    
Using,
\begin{align}
    \coth(x_{1}+ x_{2}) = \frac{\coth(x_{1})\coth(x_{2}) + 1}{\coth(x_{1})+\coth(x_{2})}
\end{align}
\begin{align}
    \implies \tilde{\Sigma}_{1 K}^{(2)}(\omega) = \bigl(\tilde{\Sigma}_{1 R}^{(2)}(\omega)-\tilde{\Sigma}_{1 A}^{(2)}(\omega)\bigr)\coth(\frac{\beta}{2}\omega)
\end{align}
So, the self-energy satisfies FDR. As a result, as proved in Appendix \ref{lblb10}, the corresponding Green's function $G_{B}$ also satsfies the FDR.
\end{widetext}

\end{document}